\documentclass[iop]{aastex62}
\usepackage{float}
\graphicspath{{./HotCopyFigs/}}
\newcommand{\HII}{H\,{\sc ii}}

\newcommand{\um}{\,$\mu$m}
\newcommand{\kms}{\,km\,s$^{-1}$}

\definecolor{bondiblue}{rgb}{0.0, .58, .71}

\shorttitle{Ionized Gas in the NGC 5253 Supernebula}
\shortauthors{Beck et al}

\begin{document}

\title{Ionized Gas in the NGC 5253 Supernebula: High Spatial and Spectral Resolution Observations with the JVLA and TEXES \\
   }

\correspondingauthor{Sara C. Beck}
\email{becksarac@gmail.com}
\affil{School of Physics and Astronomy,
Tel Aviv University, Ramat Aviv ISRAEL 69978}

\author{Sara C. Beck}
\affil{School of Physics and Astronomy,
Tel Aviv University, Ramat Aviv ISRAEL 69978}

\author{John Lacy}
\affil{Department of Astronomy,
University of Texas at Austin, Austin TX 78712} 

\author{Jean Turner}
  \affil{Department of Physics and Astronomy,
  UCLA, Los Angeles, CA 90095-1547}

  \author{Hauyu Baobab Liu}
  \affil{Academica Sinica,
  Institute of Astronomy and Astrophysics,
   Taipei 10617, Taiwan}
   
  \author{Thomas Greathouse}
\affil{Southwest Research Institute, San Antonio TX 78228-0510}

\author{S.M.Consiglio}
\affil{Department of Physics and Astronomy,
  UCLA, Los Angeles, CA 90095-1547}

\author {Paul T.P. Ho}
 \affil{Academica Sinica,
  Institute of Astronomy and Astrophysics,
   Taipei 10617, Taiwan}

\begin{abstract}

The youngest, closest and most compact embedded massive star cluster known excites the supernebula in the nearby dwarf galaxy NGC 5253.  It is a crucial target and test case for studying the birth and evolution of the most massive star clusters. 
We present observations of the ionized gas in this source with high spatial and spectral resolution.  The data includes continuum images of free-free emission with $\approx0.15$\arcsec\ resolution made with the JVLA at 15, 22 and 33 GHz, and a full data cube of the [SIV]10.5\um~ fine-structure emission line with $\approx4.5$\kms\ velocity resolution and 0.3\arcsec\ beam, obtained with  TEXES on Gemini North.  We find that 1) the ionized gas extends out from the cluster in arms or jets,  and 2) the ionized gas comprises two components offset both spatially and in velocity.  We discuss mechanisms that may have created the observed velocity field; possibilities include large-scale jets  or a subcluster falling onto the main source.  
  \end{abstract}
\section{Introduction}

 The fate of gas in young star clusters is crucial for cluster formation and evolution: stellar populations, abundances, and the
 existence of the cluster as an entity depend on the cluster's retention or expulsion of gas and subsequent evolution\citep{2017MNRAS.465.1375S}.  
  
 Cluster winds can disturb the interstellar medium on galactic scales \citep{1987AJ.....93..276H, 2013Natur.499..450B}.   But data on cluster gas kinematics is lacking because embedded massive clusters are rare (there are none at present in the Galaxy), small, distant, and highly obscured and thus challenging to observe.   The supernebula in NGC 5253 is the closest and best target for such study and the ionized gas has been extensively studied at  radio and infrared wavelengths.  

NGC 5253 is a peculiar dwarf spheroidal galaxy at  3.8 Mpc distance \citep{2004ApJ...608...42S} and the site of an intense starburst that has formed hundreds of massive star clusters in a few hundred
 Myr \citep{2015ApJ...811...75C, 2016ApJ...823...38S,2013A&A...550A..88W}.  The radio and infrared emission of the galaxy is dominated by a single remarkable supernebula, which has been extensively observed (\citet{2003Natur.423..621T}, \citet{2004ApJ...602L..85T}, \citet{2012ApJ...755...59B}, and references therein).  The embedded star cluster exciting the nebula has ionization $N_{lyc}\sim3.3\times10^{53} s^{-1}$ \citep{2015Natur.519..331T} indicating a large population of massive stars, total mass $10^{5.5}-10^6 M_\odot$, and is estimated to be only $\approx 1$ Myr old \citep{2015ApJ...811...75C}. The gas ionized by the embedded cluster is dense,  $n_e\approx3\times 10^4 cm^{-3}$, \citep{2004ApJ...602L..85T}. 
The cluster exciting the nebula has previously been identified with Cluster 11 of \citet{2015ApJ...811...75C} but improved astrometry (\citet{2018ApJ...860...47C}, Smith 2019 private comm) demonstrates that this is not the case.  The cluster exciting the nebula must be so deeply embedded \citep{1997AJ....114.1834C, 2003Natur.423..621T, 2004ApJ...612..222A} as to be undetectable in visible and near-infrared light. 

The supernebula is small, with a core size of $\lesssim 0.2$\arcsec\ \citep{2004ApJ...602L..85T}, immersed in 
fainter extended H$\alpha$ and radio emission \citep{1997AJ....114.1834C,1998AJ....116.1212T}. \citet{2007ApJ...670..295R} mapped the $H53\alpha$ recombination line in NGC 5253 with the VLA, using 
  a 2\arcsec\ beam and 44\kms\ velocity resolution, and  found a gaussian line profile of FWHM $58\pm12$\kms\ and an apparent velocity gradient of  10\kms\ \arcsec\ $^{-1}~$ increasing from the NW to the SE, which they suggest shows rotation of the supernebula.   
 Infrared  HI recombination lines were observed 
 by 
 \citet{2018ApJ...860...47C}, who mapped $Br\alpha$ 4.05\um~with $\approx0.1$\arcsec\ resolution and found a line profile combining a strong core 65-75\kms\ wide  and a weak plateau with FWHM $\approx150$\kms, which they suggest arises in stellar winds; they also observed a velocity gradient north-south across the supernebula itself.  \citet{2012ApJ...755...59B} measured the infrared line of [SIV] 10.5\um\ with TEXES at the NASA IRTF with $\approx4.5$\kms\ total resolution (instrumental and thermal) and  1.4\arcsec\ spatial resolution.  They found no spatial dependance, and an asymmetric line profile with excess emission at blue velocities; they modelled the [SIV] line as having a main component centered at 391\kms, consistent with the galactic velocity,  and another gaussian component at 368\kms, and suggested that the  gas may be flowing out of the dense ionized region.  However, the spatial resolution and signal-to-noise of those observations were not sufficient to support further 
modelling.  Observations of moleular gas in this region \citep{2017ApJ...846...73T, 2017ApJ...850...54C} found that the cluster coincides with a very compact molecular cloud and the cloud's CO line profiles are smooth, gaussian and narrow in velocity, with FWHM $\approx$22\kms\  . It is suggested that this cloud is actually inside the cluster and is composed of many small, hot and dense molecular cores. 

Clearly the  gas kinematics  of the  NGC~5253 supernebula are complex and the source still not understood.  This motivated us to observe the ionized gas in the NGC 5253 supernebula with the highest attainable combinations of spatial and spectral resolution.  We report here on [SIV] data obtained on the Gemini North telescope with true velocity resolution $\approx4.5$\kms\ and spatial resolution $\approx0.3$\arcsec\ and on 9 mm, 1.5 cm and 2 cm maps made with the Karl G. Jansky Very Large Array (VLA) at spatial resolution $\approx0.15$\arcsec. These data have sensitivity, and for the [SIV] spatial resolution, significantly better than the earlier results  of \citet{2012ApJ...755...59B} and \citet{2004ApJ...602L..85T}.   In these data we have the clearest picture to date of sub-structure in the main cluster, sufficient to locate  independant velocity components of the ionized gas. 

\section{Observations and Data Reduction}
\subsection{Radio Continuum Maps}
NGC 5253 was imaged at 33, 22 and 15 GHz with the JVLA in the high resolution A array as part of project 16B-067. The data went through the standard JVLA pipeline process and were  calibrated, cleaned, self-calibrated and imaged with CASA and AIPS.  The 22 GHz data had phase problems that did not permit self-calibration and are not presented here.  Observational parameters are shown in Table 1. The peak and total fluxes were 17 mJy/bm, 25 mJy at 33 GHz and 9.7mJy/bm, 17 mJy at 15 GHz.  Mutiple sub-structures in the supernebula  are apparent in these maps and will be discussed below.  For  reference, we show in the Appendix an image with the  sub-structures marked. 
\subsubsection{Mid-Infrared Observations}
TEXES, the Texas Echelon Cross Echelle Spectrograph \citep{2002PASP..114..153L}, operates between 4.5 and 25 \um\ and has spectral resolution $R\approx 4000-100,000$.  The mid-infrared spectrum of NGC 5253 is dominated by the high excitation lines of [NeIII] (15.5\um) and [SIV] (10.5\um); [NeIII] cannot be observed from the ground so [SIV] is the preferred probe of ionized gas kinematics.  [SIV] has two advantages over HI recombination lines.  First, the extinction.  $A_{[SIV]}$ (the extinction at 10.5\um~), while certainly much less than $A_v$,  cannot be well determined because it depends on the shape and depth of the 9.7\um~ silicate absorption.  The silicate feature is weak in NGC 5253 \citep{1999MNRAS.304..654C} and we follow \citet{2005A&A...429..449M}, \citet{2016ApJS..224...23X} and \citet{1999MNRAS.304..654C} that $A_{[SIV]}$ will be significantly less than $A_{Br\gamma}$ and comparable to $A_{Br\alpha}$.  Second, [SIV] is much less affected by thermal broadening than any hydrogen line can be.  Thermal broadening in an \HII~ region at $T_e\sim10^4 K$ gives any hydrogen line FWHM about 20\kms\ ;  at the same temperature a line of a metal ion of mass $m_i$ will have width $\sqrt{\frac{m_H}{m_i}}\times$ that of  hydrogen.  So the thermal broadening for $S^{+++}$ is only $\approx 3.5$\kms , which adds in quadrature to the instrumental resolution for true resolution of $\approx4.5$\kms\ . 

NGC 5253 was observed at Gemini North on 15 March 2017 under GN-2017A-Q-57, using TEXES in high resolution mode, the 32 line/mm echelle and a 0.5\arcsec\ wide, 4\arcsec\ long slit.   The slit was aligned north-south and stepped east-west across the targets. The scans are combined into data cubes for each order of the echelon; the cubes for one order has dimensions  22 pixels R.A. $\times$ 27 in Declination $\times$ 256 in velocity and each pixel is 0.29\arcsec  $\times$ 0.14\arcsec $\times$ 0.95\kms\ .   The asteroid Daphne was the calibrator and the point-spread function was measured on $\epsilon$Ori.  Because the source is so highly obscured, the targets were acquired by sending the telescope to the coordinates of the radio continuum peak. The position of the [SIV] peak agrees to within 1 pixel with the radio peak.

 \section{ Spatial Structure Of the Ionized Gas}
 \subsection{Radio Continuum}
\begin{figure}[h]$
\begin{array}{cc}
\includegraphics*[scale=0.5]{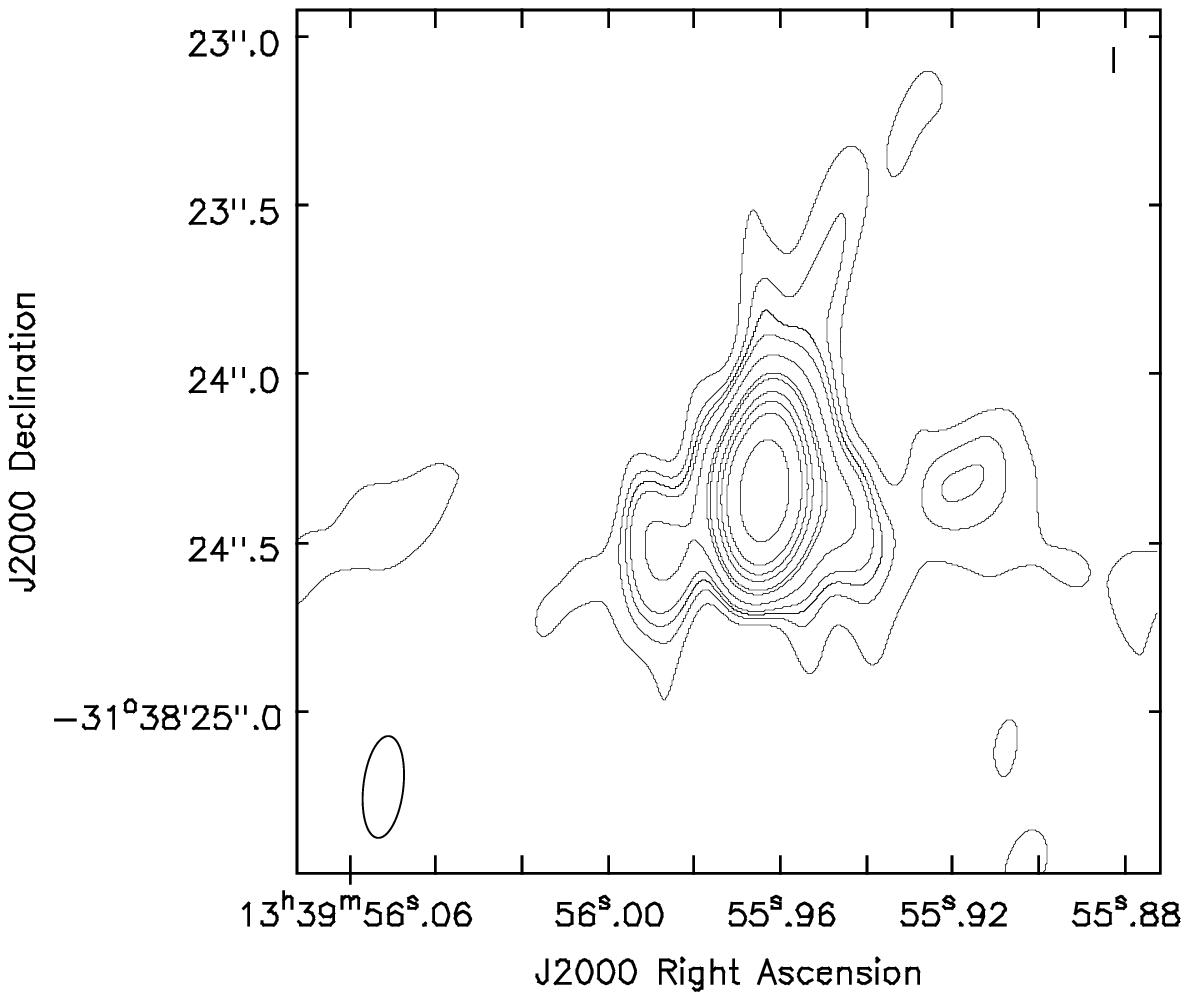}  &  \includegraphics*[scale=0.5]{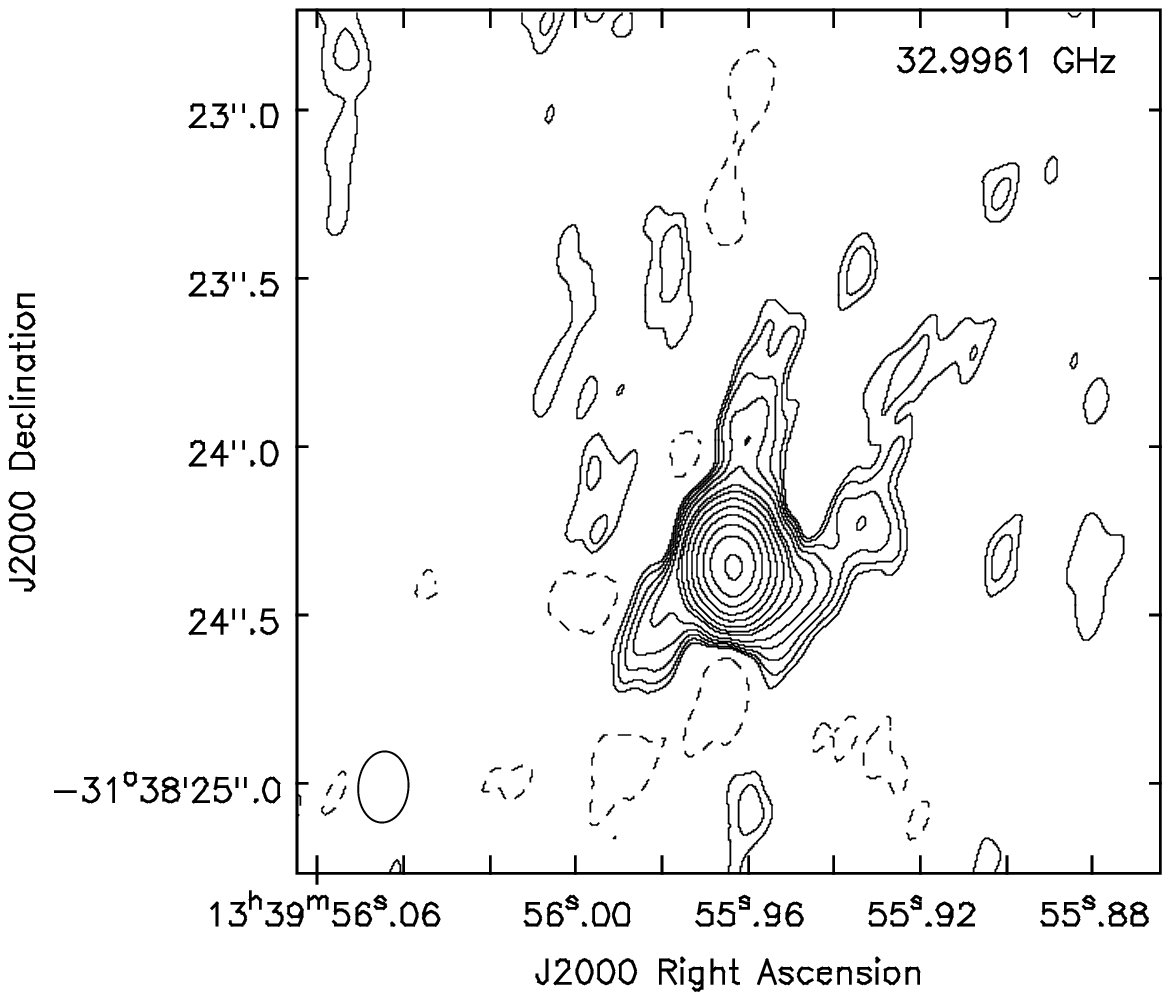}  \\
a) & b) \\
\end{array}$
\caption{a) The radio continuum emission of NGC 5253 at 15 GHz. b) The radio continuum emission at 33 GHz.    Both maps are self-calibrated and Briggs weighted. 
Contours are at  $2^{n/2}\times$ intervals of 0.1 mJy/bm at 15 GHz and 0.3 mJy/bm at 33 GHz; dotted contours are negative.  
Beams are shown in lower left. }
\end{figure}
Briggs-weighted (robustness 0) maps of the radio continuum at 15 GHz (U band) and 33 GHz (Ka band)  are shown in  Figure 1. The Ka and U maps agree remarkably well considering their different spatial resolution. There is a very strong and small main source; in the highest resolution (uniform weighted) Ka map, which has a $0.13\times0.03$\arcsec\ beam, the deconvolved source size  is $0.14\times0.03$\arcsec ,  about  $2.6\times0.8$ pc at 3.8 Mpc.  The radio maps show distinct resolved structures; we refer to them with the labels they have on the figure in the Appendix. 
There are extensions or 'arms' extending north and west, which we call the N and W arms,  and a short protusion we call the  'SE spur' in the south-east direction. The SE spur also appears in the echellograms of Br~$\alpha$ \citet{2018ApJ...860...47C} made
with long slits oriented east-west.  
  The `arms'  and 'spur' are  
at least 10$\sigma$ with peak fluxes of almost 1 mJy/beam and appear at both Ka and U radio bands with 
similar shape and extent.  
The N and W arms may have been hinted in previous K band images \citep{2000ApJ...532L.109T} and in the 7mm  maps of \citet{2004ApJ...602L..85T}, which had noise at least a factor 5 higher than these. The 7mm maps also show a filament or arc south of the main source. This filament is not seen in Figure 1b, perhaps because that region is undersampled, as shown in the negative contours. 
The weak 1.3 cm source  $\sim$0\farcs5 due east of the supernebula, which was seen by \citet{2000ApJ...532L.109T} and is often referred
to as the other component of a ``double cluster",  does not appear in the radio. 

\subsection {[SIV] Spatial Distribution}
  \begin{figure}[h]
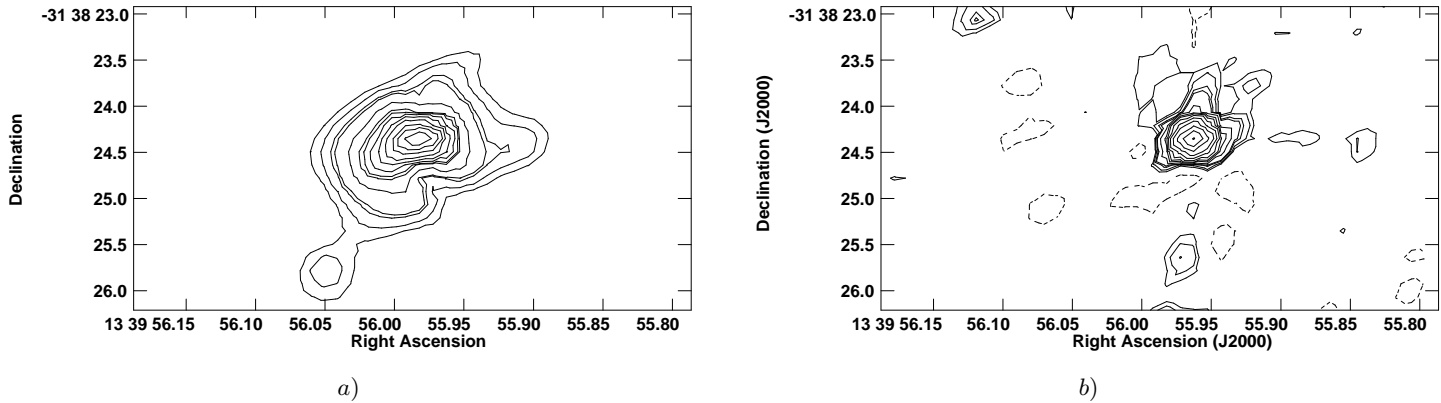
$
 \begin{array}{cc}
\includegraphics*[scale=0.5]{FIGURE1.EPS} & \hskip 0.2cm \includegraphics[scale=0.5]{KASCFOO.EPS} \\
a) & b) \\
\end{array}$
\caption{a) The zeroth moment (total intensity) of the primary [SIV] data cube of NGC 5253, sharpened to correct for the PSF.  This image includes velocities  278-508\kms\.  Contour levels are  $0.018\times2^{n/2},  n=0,1,2..$ in units of $erg(s~ cm^{-2} cm^{-1})^{-1}$.  b)  The Ka radio continuum convolved to the beam and matched to the grid of the [SIV] map; contours are $2^{n/2}\times 0.3$mJy/bm( $5\sigma$) and dotted contours are negative. }
\end{figure}
 The TEXES data presented here have significantly higher spatial resolution and dynamic range than the earlier [SIV]  
 measurements.  Figure 2 shows an image of the integrated line intensity which has been sharpened by deconvolving the point spread function (the point-spread function was not very stable and was influenced by seeing and telescope jitter, so this image is more properly considered a sharpening rather than a true deconvolution).  To produce this image the data were re-binned into 6.54\kms\ pixels and further smoothed by 2 points  and the noisy ends of the order truncated so as to work only with high S/N; velocities of 278-508\kms\ were included.  The resulting map resembles the radio continuum structure, indicating that the 10.5\um~ extinction is low or moderate; this agrees with \citet{2003Natur.423..621T}'s estimate of $A_{Br\alpha}\sim0.5$mag.  As in the radio the main source is resolved and extended NW-SE and there is an extension due W and another slightly W of N.   Figure 2b shows the Ka map convolved to the resolution of the [SIV], and comparing it with the [SIV] map of Figure 2a shows that the radio emission north of the main source, which appears as multiple arms in the self-calibrated radio maps, will all fall in the NW feature of the [SIV], and that the western extension of the radio and [SIV] agree.  
     
The [SIV] emission extends south and east of the main source,  in the same direction as, but further than, the radio emission 'spur' .   The secondary [SIV] emission region further south-east of the main source (r.a.13h39m56.05s, dec -31d38m25.75s) coincides with a weak radio peak seen in archival B-array maps but apparently resolved out by the current observations.   The small clump of radio emission due south of the main source has the ionization of $\approx 150$ O7 stars but no optical or [SIV] emission; this can be explained by the high extinction expected from its position on a molecular clump in the CO streamer.  There is no [SIV] emission towards the secondary K-band peak seen by \citet{2018ApJ...860...47C} $\approx0.4$\arcsec\ due east of the main source.

\begin{deluxetable}{ccccc}
\tablecaption{Observational Parameters}
\tablenum{1}
\tablehead{\colhead{Date} & \colhead{Telescope} & \colhead{Wavelength } & \colhead{Beam Size$^a$} & \colhead{RMS } \\ 
\colhead{} & \colhead{} & \colhead{} & \colhead{} & \colhead{} } 
\startdata
18//12/16  &  VLA  &  33 GHz  &  $0.21\times0.149$\arcsec  &  $7\times10^{-5} Jy/bm$ \\ 
19/11/16  &  VLA  & 15 GHz  &   $0.3\times0.11$\arcsec &   $1.2\times10^{-5} Jy/bm$ \\ 
 21/10/16  &   VLA  &  22 GHz  &  $0.21\times0.089$\arcsec  &   $2.1\times10^{-5} Jy/bm$ \\ 
15/3/17  &  TEXES-Gemini North   &  10.5\um\  &  $0.3\times0.3$\arcsec  &  $5\times10^{-3}~erg(s~cm^2~sr~cm^{-1})^{-1}$ \\ 
\enddata
 \tablenotetext{a}{Beam sizes of radio maps from Briggs-weighted, and at 15 and 33 GHz self-calibrated,  maps. }
\end{deluxetable}
 \subsection{Ionized Gas Distribution from Radio Continuum and [SIV]10.5$\mu$m Emission Compared to Optical and to Molecular Gas}
 
  The dense, bright ionized gas imaged in the radio continuum and [SIV] shows the most intense concentration of very young stars. How does the distribution of radio and IR nebular line emission compare to that of the molecular gas from which the cluster formed, and the more dilute, less obscured regions of earlier star formation?  

 Figure 3 shows the [SIV] emission and the radio continuum overlaid on a Hubble Legacy image of emission in the 814nm wide filter; the images are registered 
 following \citet{2018ApJ...860...47C}, which agrees with GAIA astrometry (Smith, private communication).  The stellar and nebular emission at 814nm appears as two regions of disturbed and complex structures, separated by an obscured strip.  The [SIV] and radio emission are both centered on the region of high obscuration.  The radio 'arm' N of the center agrees with a bright arc in the 814nm and the 'arm' W of the core with a patch of obscuration.   The N and W extensions of the [SIV] emission, which follow the radio 'arms', coincide with the same optical features.  On the southeast the
   radio 'spur' overlaps the brightest region in the 814nm field, the optical continuum source  ``Cluster 5" \citep[][; Fig. 3]{2015ApJ...811...75C} .  The [SIV] emission extends further to the southeast than does the radio, overlapping the whole region of 814nm emission, and the secondary [SIV] source lies in the obscured region south-east of ''Cluster 5".  The greater extent of [SIV] emission southeast of the peak, and its agreement with the bright 814nm emission, suggest that obscuration is relatively low in that region.  
   
  Figure 4 compares the ionized gas to the molecular by overlaying the radio continuum on the CO(3-2) emission mapped by  \citet{2017ApJ...846...73T} and shows that the core of the embedded HII region coincides with the CO cloud D1 \citep{2017ApJ...846...73T,2017ApJ...850...54C} .  This is consistent with the high optical extinction at that position seen in Figure 3.   The radio continuum however avoids the large Cloud D4 southwest of the supernebula; since the radio is unaffected by extinction this implies that the \HII~ region is ionization-bounded toward the southwest.  This is in contrast to the N and W radio 'arms'  and their counterparts in the nebular emission northwest of the supernebula.    The southern 'spur' of radio emission coincides with and resembles a CO 'spur'  on Cloud D1.
  
   \begin{figure}
 $\begin{array}{cc}
\includegraphics[height=3.5in, width=3.5in]{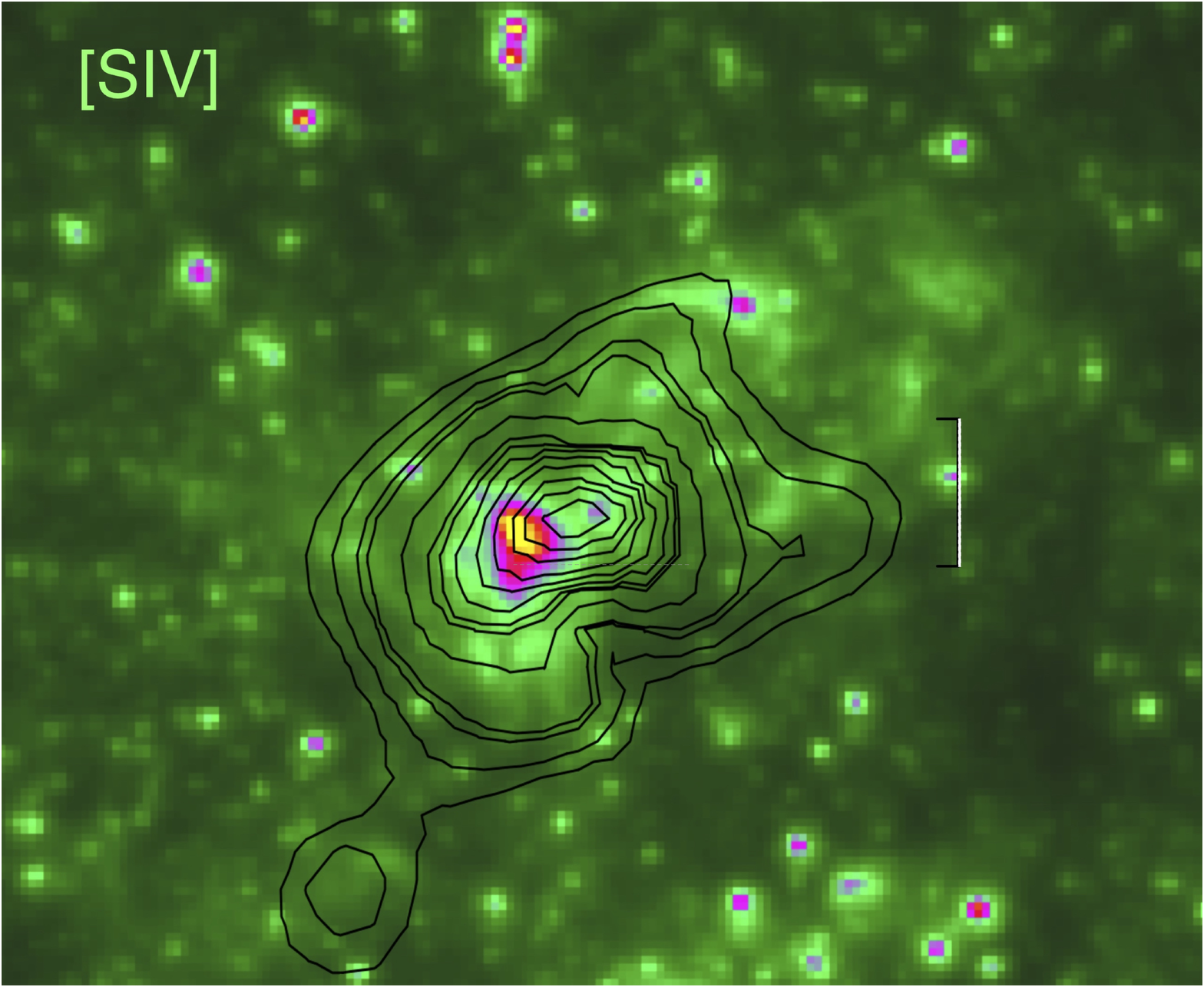}  & \includegraphics[height=3.5in, width=3.5in]{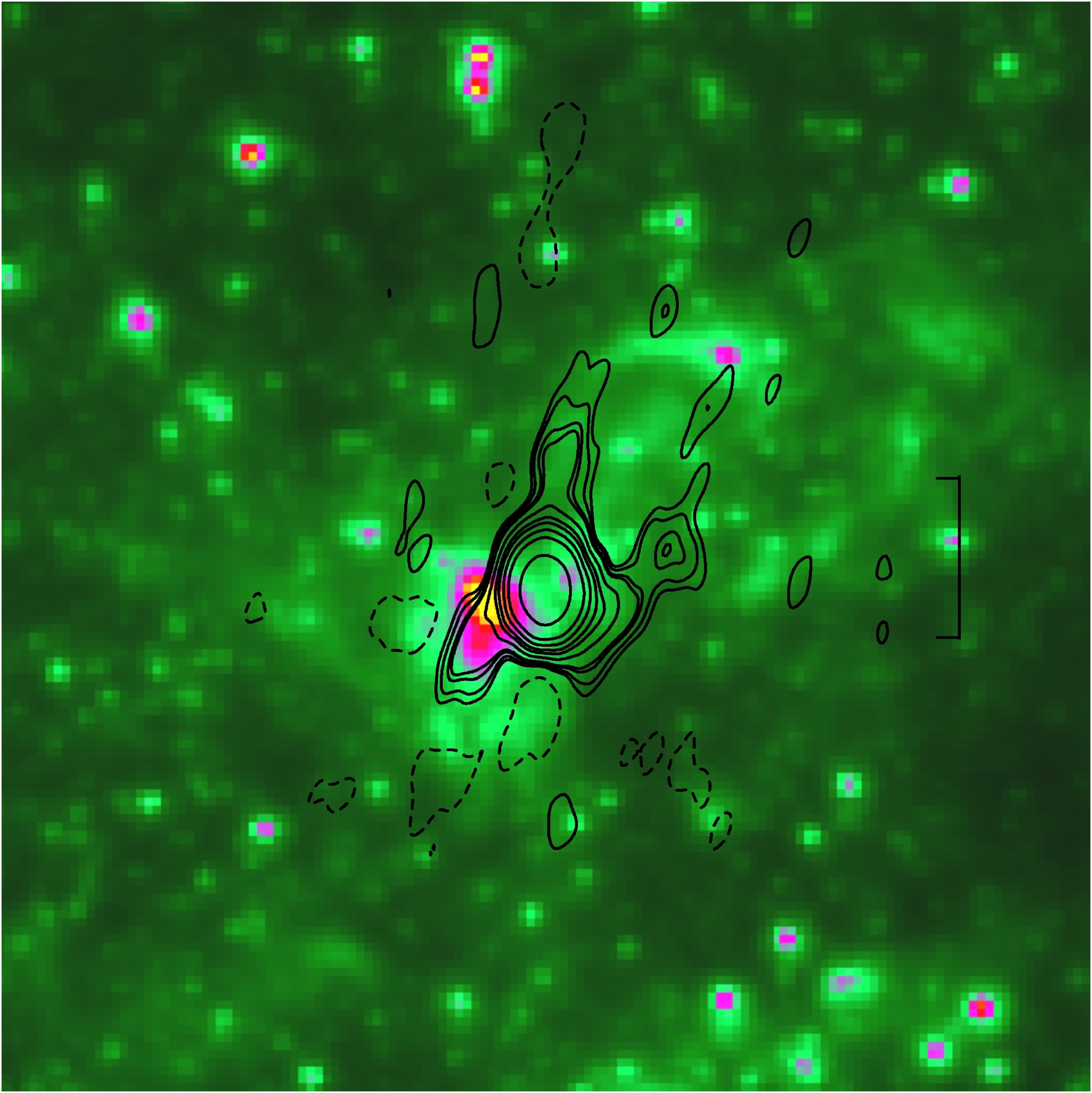} \\
a) & b) \\
\end{array}$
\caption{a): The zeroth moment of the [SIV] data cube, sharpened, in contours overlaid on an 814nm image from the Hubble Legacy Archive. Contour levels are as in Figure 2. North is up, East left, and the  black and white vertical bars are 1\arcsec~.  b): The 33 Ghz emission with contour levels as in Figure 1, overlaid on the 814nm image. The 814nm image was registered by alignment with cataloged stars in the field. }
\end{figure}

 \begin{figure}
 \begin{center}
\includegraphics[scale=0.2]{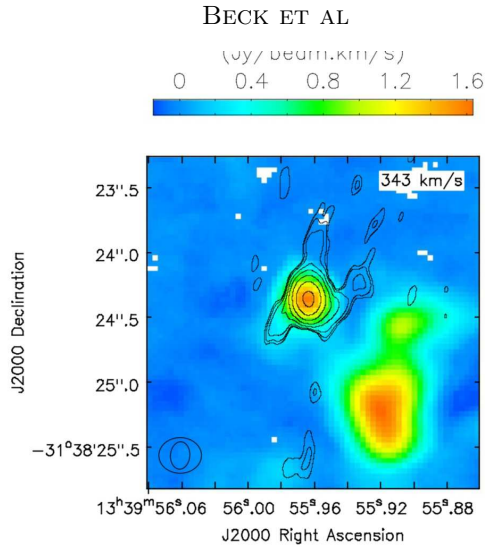}
\end{center}
\caption{Contours of 33 GHz continuum are overlaid on the CO(3-2) integrated intensity from \citet{2017ApJ...850...54C} 
and \citet{2017ApJ...846...73T}.  
The  CO intensity is given by the color wedge and the radio contours are $0.1,0.2, 0.4,1, 2,4,8,16\times 1.13$ mJy/bm, starting at 2 mJy/bm. }
\end{figure}

 
\section{Kinematics of the Ionized Gas}
The maps presented above show complex spatial structures in the ionized gas.   We now examine the motions of gas in the NGC 5235 supernebula with the aim 
of arriving at a working three-dimensional model of the region and some insight into its history and evolution. The first step is to determine the underlying velocity field of the host galaxy.   NGC~5253 is a dwarf spheroidal with very little
 rotation \citep{1989ApJ...338..789C}; the usually accepted galactic velocity is 391\kms\ and stellar features give 389\kms\ \citep{2004ApJ...610..201S}.  Optical features such as $H\alpha$ have very complicated 'bubbly' \citep{1982MNRAS.201..661A} velocity fields in the center of NGC~5253 because the gas has been disturbed by the many central clusters; better comparisions to the [SIV] are  tracers likely to be concentrated on and coincident with the supernebula and embedded cluster.  Among them the molecular cloud coincident with the supernebula is at 387\kms~\citep{2017ApJ...846...73T}, the $Br\alpha$ line in the central 2\arcsec\ at 391\kms\ \citep{2018ApJ...860...47C} and the $H30\alpha$ line in the inner 0.5\arcsec\ at 389.6\kms\ (Turner, private communication 2019).  In this paper we will take 391$\pm3$\kms\ to be consistent with 'galactic velocity'; the estimated error reflects both resolution and the accuracy with which the infrared wavelength scale can be registered.
  
  Figure 5 displays the line profile in the sharpened [SIV] image at every other spatial pixel. Because of the large dynamic range of the data and the importance of velocity features other than the peak, we show two intensity scales: the full scale in Figure 5b shows the center of the emission region and 5a shows  the lower levels of emission away from the peak.  
 
\subsection{The Main Peak, Pedestal, and Continuum Features}

 \begin{figure}
\includegraphics[height=4.5in,width=4.5in]{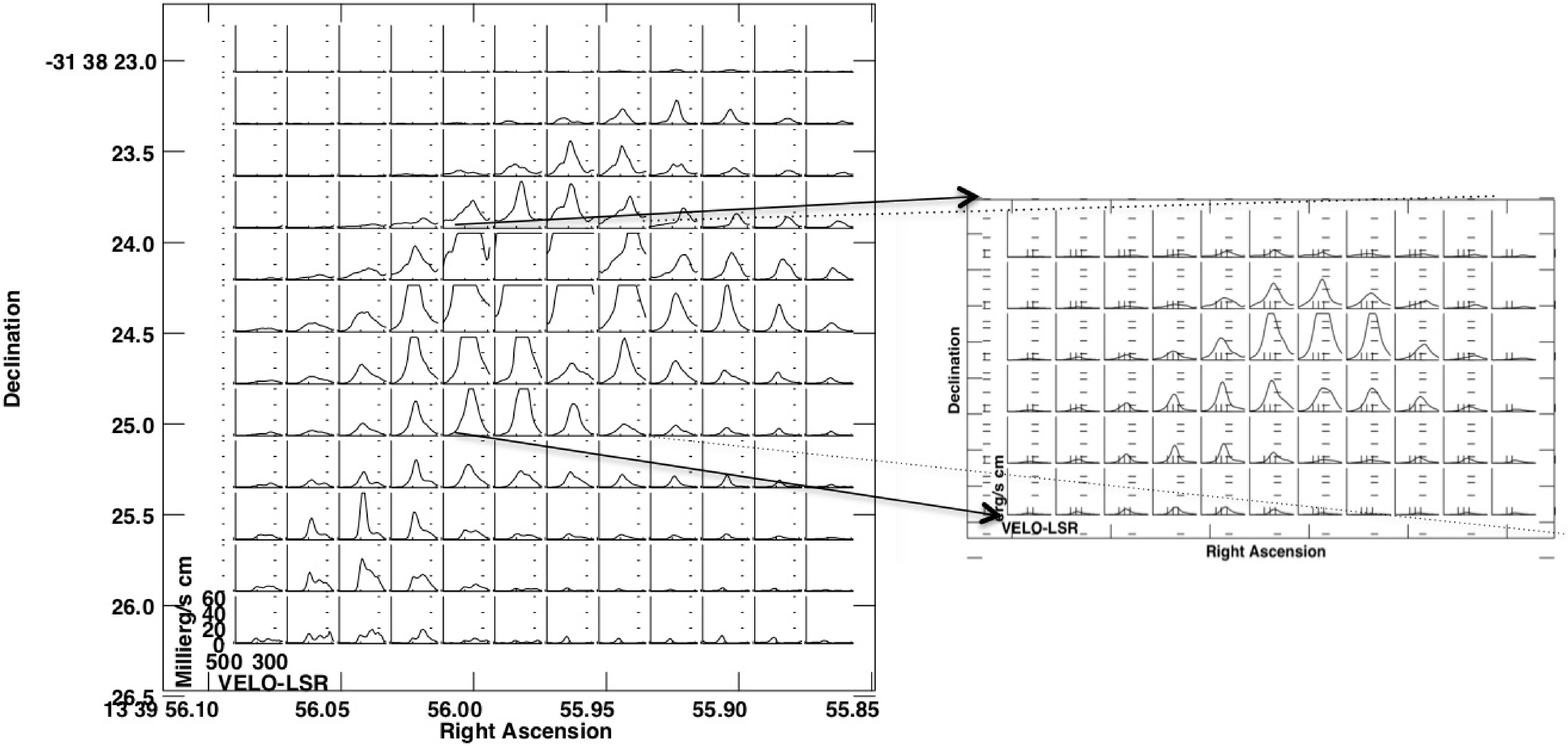}
\caption{The [SIV] line profile in the $0.14\times0.14$ \arcsec\ pixels of the sharpened data cube.  The left figure covers the entire emission region and shows every pixel in R.A. and Declination; the right shows every pixel over the core of the source. The units are, in full, $ergs(s~cm^{-2}cm^{-1})^{-1}$ and different intensity scales are used in order to display the full range of emission: the maximum value is $60\times10^{-3} ergs(s~cm^{-2}cm^{-1})^{-1}$  in the left figure and $2~ergs(s~cm^{-2}cm^{-1})^{-1}$  on the right.  }
\end{figure}

 Examining the [SIV] data cube reveals several gas components distinct in space as well as velocity.  Even with the 0\farcs 3 
 spatial resolution of TEXES on Gemini most beam positions include more than one of these components, which creates a rich and confusing variety of spectral line profiles.   For this reason we believe the first moment (intensity-weighted velocity) of the data cube to be deceptive; while such a moment map would show apparent overall trends of velocity,  these may well arise from  distinct velocity components  appearing at different strengths in different positions, rather than from an organized dependence of velocity on position.  The ionized gas 'jets' seen at optical wavelengths \citep{1982MNRAS.201..661A}
 and the radio and infrared 'arms' demonstrate that the gas in this region holds complex spatial structures; are they reflected in the velocities, and if so, how? 
  
We first review the line profile of the main source taken as a whole, in Figure 6a.  This shows the unbinned and unsmoothed profile in a 0.5\arcsec\ diameter beam on  the main peak, including $\approx80$\% of the total [SIV], the
entire supernebula and CO Cloud D1. The best single-gaussian fit, shown in Figure 6a, does not fit either the peak or the blue side of the line 
well. Its FWHM of $66\pm1$\kms~is similar to the Brackett $\alpha$ width for the central component, 
not including weak line wings, found by \citet{2018ApJ...860...47C}.    
 Figure 6b demonstrates that the sum of two gaussians, one  
centered at the Brackett $\alpha$  
velocity of $391\pm1$\kms\ \citep{2018ApJ...860...47C} and a blue feature at $350\pm3$\kms~, is much closer to the profile.  This confirms with high S/N \citet{2012ApJ...755...59B}'s finding of bulk non-turbulent gas motions blue of the galactic velocity.

 What are the extreme gas velocities in NGC 5253?  Figure 6a shows that the [SIV] line on the peak does not fall to  zero intensity until  at  least $\pm100$\kms\ from the peak. Combining several of the simultaneously-observed echelon orders gives an extended velocity baseline.  The spectrum in a 1\arcsec\ diameter beam on the emission peak is shown for the full range of -500 to 1150 \kms\ in Figure 6c.   There is continuum emssion at the level of $0.01~erg(s~cm^2~sr~cm^{-1})^{-1}$, 
 $\approx2\sigma$.   NGC 5253 is an unusually strong continuum source to be detected at all at such high resolution; the level observed is consistent with the 1.3 Jy seen by Spitzer in a much larger beam.  At  velocities $\pm 200$\kms\ of the peak there is evidence of a weak pedestal of emission about $1\sigma$ above the overall continuum level.   This emission pedestal is consistent with the $\approx150$\kms ~FWHM feature \citet{2018ApJ...860...47C} detect on the $Br\alpha$ line and which they suggest is a weak cluster wind.  That the red side of the pedestal is weaker is probably because of obscuration to the far side of the nebula; extinction at 10.5\um~ is not negligible. 

 \begin{figure}
\begin{center}
$\begin{array}{ccc}
\includegraphics*[height=2.5in, width=2in]{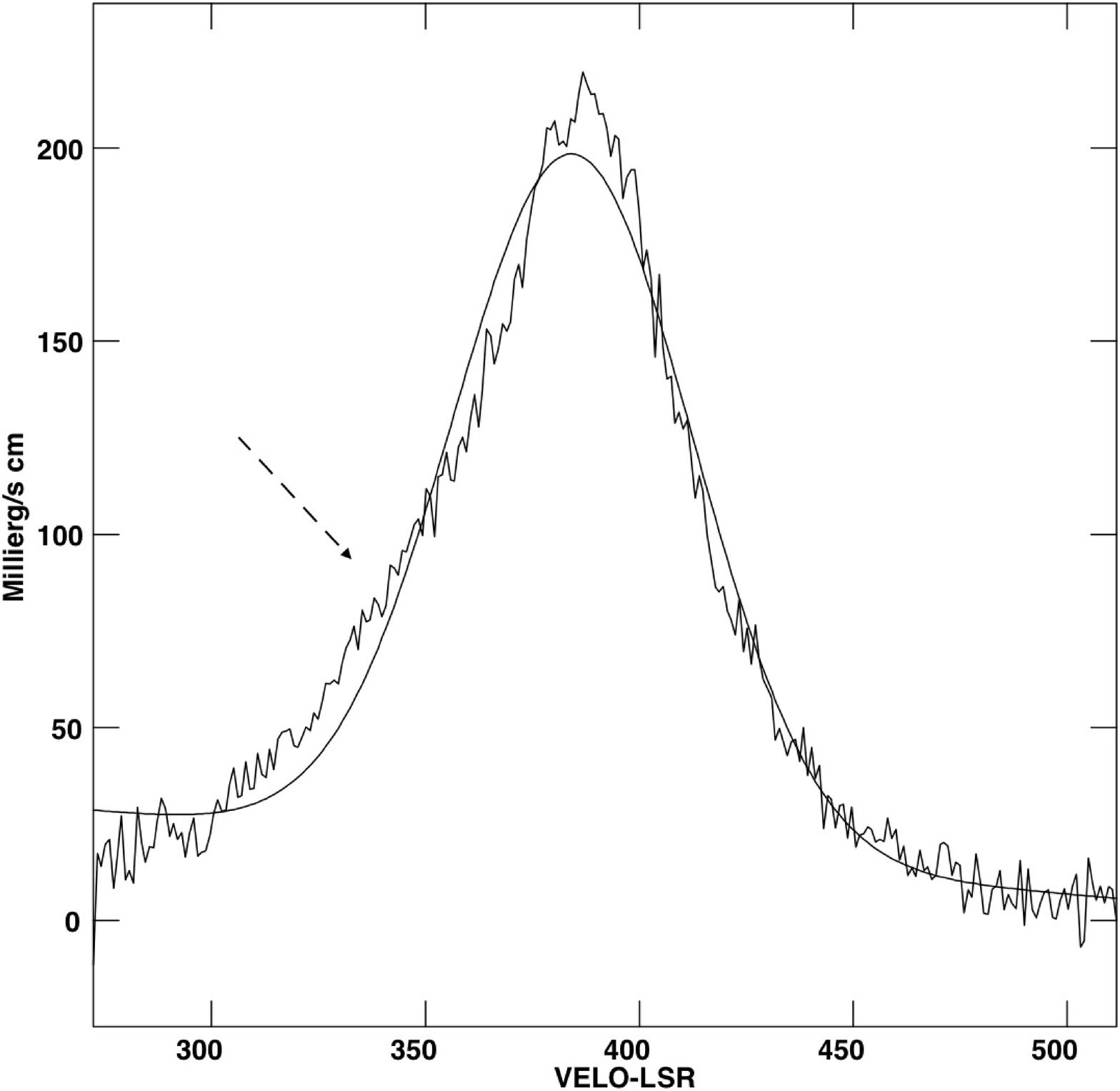} &  \includegraphics[height=2.1in, width=1.8in]{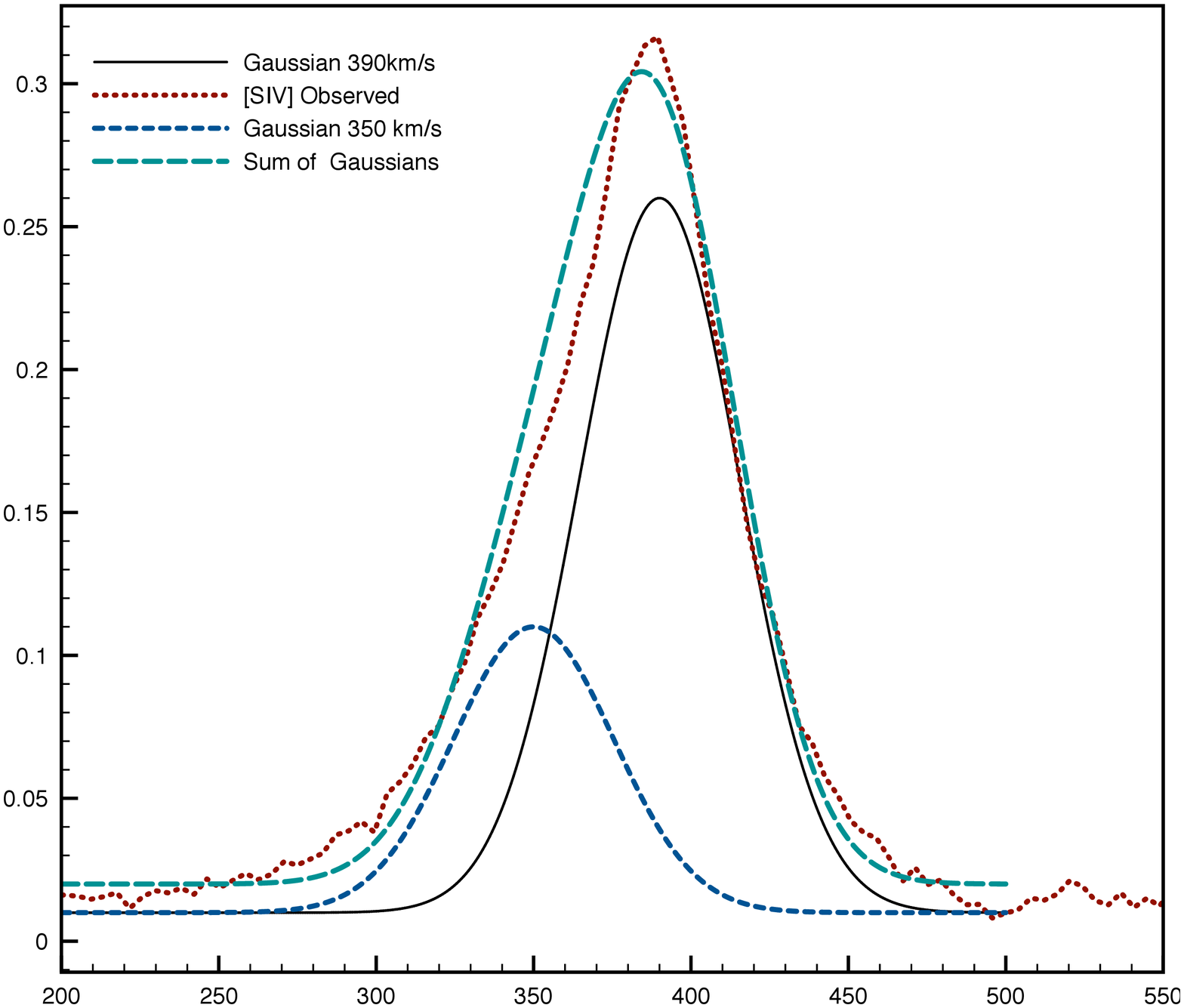} &   \includegraphics[height=2.5in,width=2in]{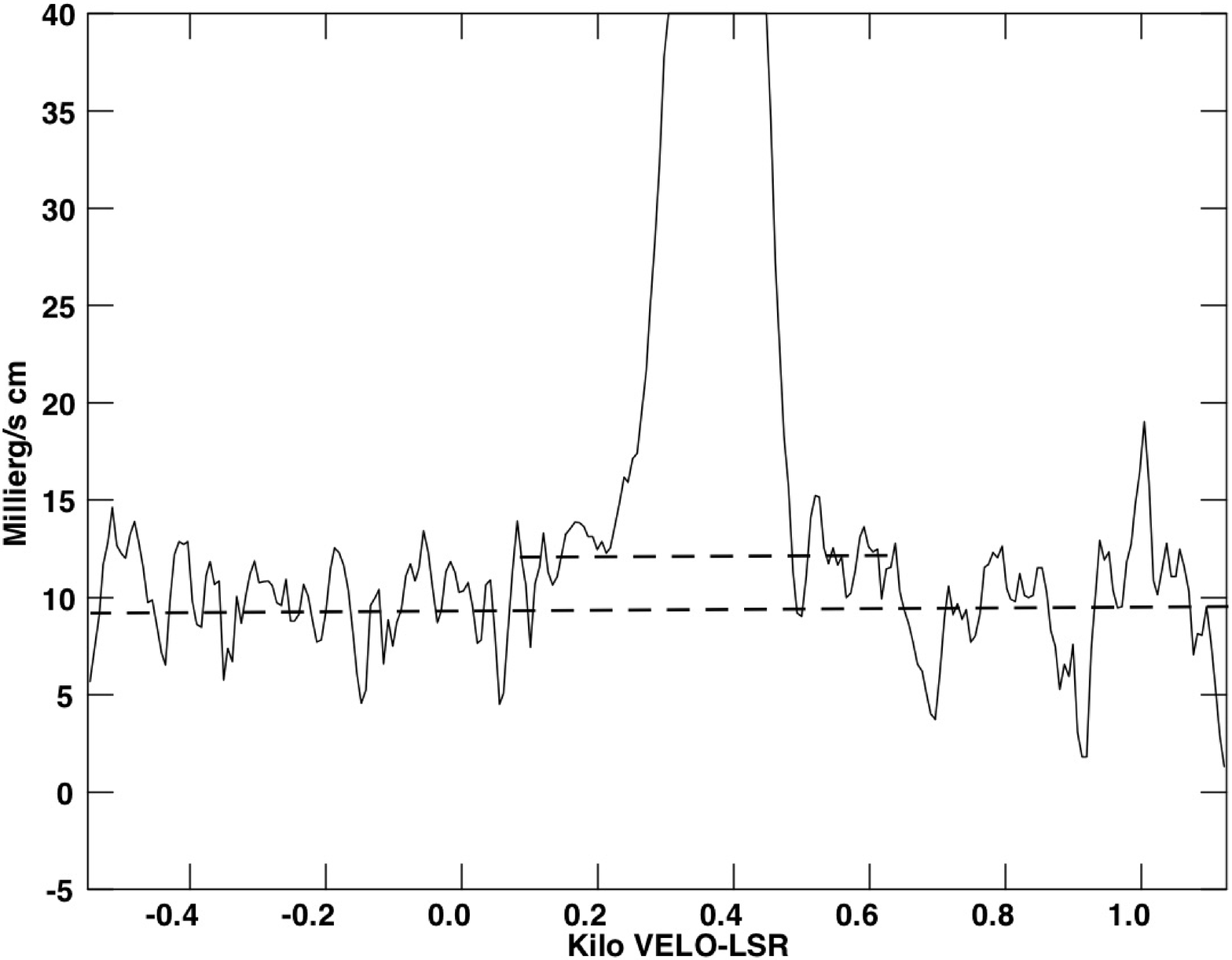}\\
a) & b) & c) \\
\end{array}$
\end{center}
\caption{a) The line profile from unsmoothed, unbinned data, taken with a 0.5\arcsec\ beam on the peak of the emission, and the best single gaussian fit thereto.  The velocity scale is in \kms\ .  The sloping baseline is an artifact caused by forcing a single gaussian fit onto a line profile with significant excess blue emission; the dotted arrow points to the spectral region most affected by the blue excess.   b) Two gaussians, representing the main velocity component and the blue-shifted feature, and their sum are shown and compared to the observed line profile. The gaussians were shifted vertically to match the continuum level of the data. c) The lowest level of emission over an extended velocity range. The velocity scale is $10^3$\kms\ .  The continuum level at $\approx0.009$ and the pedestal or blue wing  at $\approx0.013$ are marked with dashed lines.  The beam is 1\arcsec\ diameter, placed on the peak,  and the data has been binned by 7 pixels or 6.4\kms\ .}
\end{figure}

\subsection{The Blue Bump} 
The most striking non-gaussian feature of the line profiles 
is a spectrally and spatially distinct component blue-shifted of the main peak.  This blue component is clearly seen in the spectral profiles of Figure 7.   The blue feature is detected in isolation, without the main peak, in the declination range -31:38:23.7 to -31:38:24.2, which is marked in the Appendix as 'blue emission' and which largely coincides with the area of weak emisison between the NW and W 'arms'.  That line profile is shown in Figure 8a as the 'blue bump'.  It is centered at $\approx360$\kms\ and is $\approx50$\kms\ wide, overlapping considerably the main velocity peak and producing the blue wing originally reported by \citet{2012ApJ...755...59B}.That it can be distinguished in the [SIV] line profiles but was not seen as a separate feature in infrared or radio HI recombination lines is probably the effect of thermal broadening on the HI lines.   

The blue feature is $\approx20$\% as strong as the main velocity component.  With the current spatial resolution we cannot determine the spatial extent of the blue emission or to fix its center position, except to say that it is strongest in the west and NW of the supernebula and much weaker, although still present, in the east.  Figure 8b compares the (normalized) profile on the western edge of the source, minus the blue bump, to the line profile on the east. The line profiles resemble each other, although they do not perfectly match.  Better data on the weak and noisy blue feature would permit us to further disentangle the velocity components in this supernebula.    

The discovery of a spatially distinct blue feature leads to reconsideration of the previous results   
that reported a NW-SE velocity gradient interpreted as rotation \citep{2007ApJ...670..295R}) .  
  The current [SIV] data show a line peak shifting  with position, but {\it not} due to rotation.  There is in this data no evidence that the source is rotating.   Possible explanations and models of the blue bump and its role in the cluster will be discussed below.

\begin{figure}[ht]
\includegraphics[scale = 0.6]{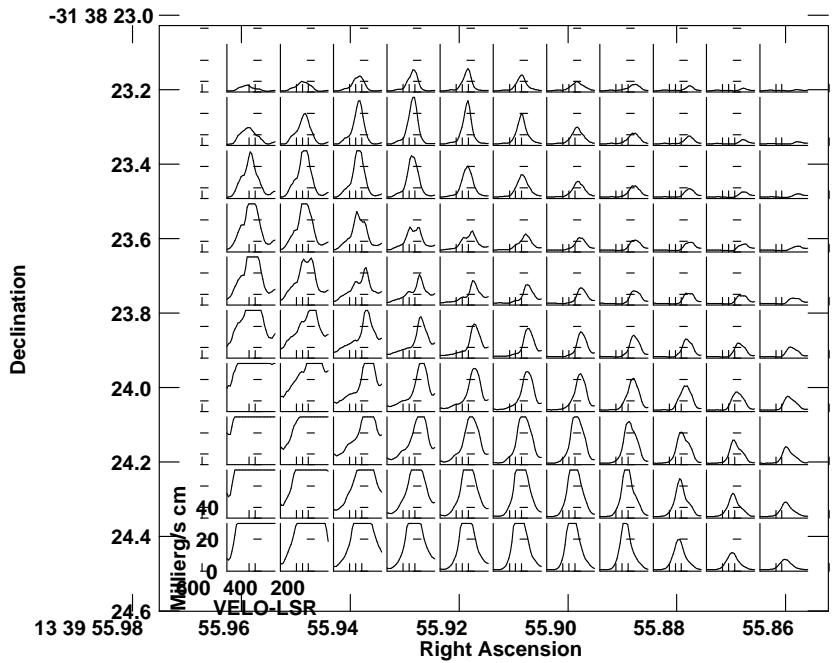}
\caption{ The spectral profile of the [SIV] line in the NW quadrant of the field, in the sharpened image, in every $0.14\times0.14$\arcsec\  pixel.   The weak blue feature is seen in isolation in the north-west. The intensity scale is clipped to display the weaker features. }
\end{figure}

  \begin{figure}[ht]$
  \begin{array}{cc}
\includegraphics[scale=0.34]{BBFIT.EPS} & \includegraphics[scale = 0.37]{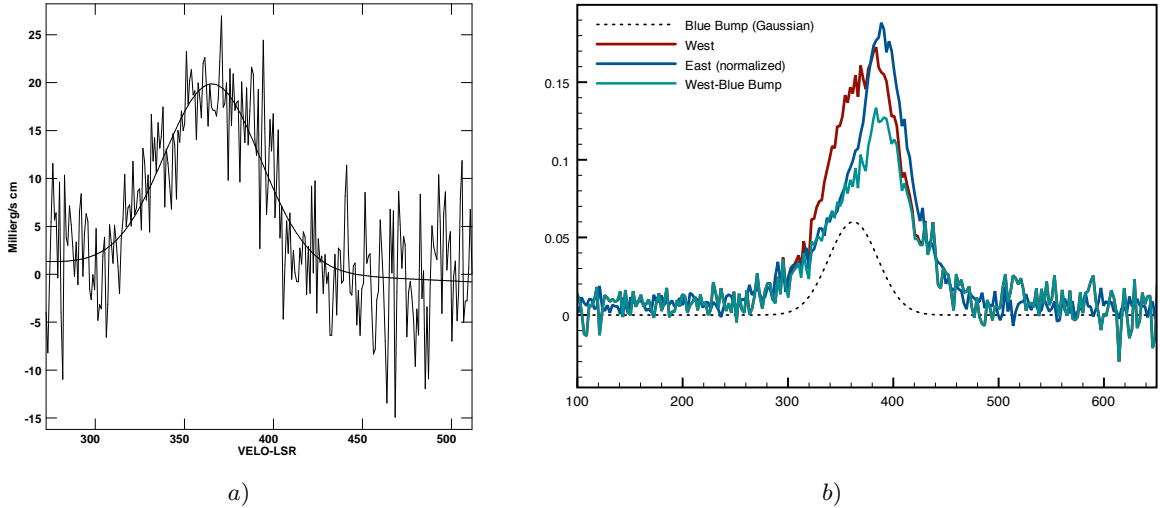} \\
a) &  b)  \\
\end{array}$ 
\caption{a) The 'blue bump' line profile from unsharpened and unsmoothed data,  in a 0.35\arcsec\ beam at the position given in the text. b) The line profile in unsharpened and unsmoothed data over the Western and Eastern halves of the source, the gaussian fit to the blue bump feature and the W spectrum minus the gaussian, as labeled on the figure. Velocity scale is in \kms\ and the intensity scale is normalized to an arbitrary value. }
\end{figure}

\subsubsection{The South-Eastern Source } 
\begin{figure}[h]
\includegraphics[width=0.5\columnwidth]{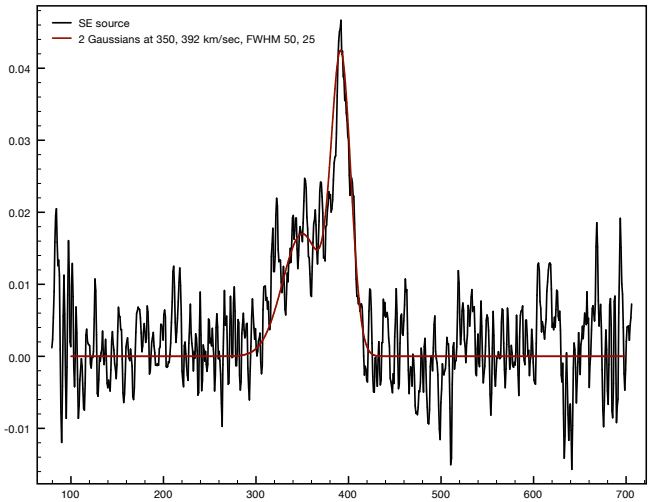} 
\caption{The [SIV] line profile in a 0.64\arcsec\ diameter circle over the SE source.  The best two-gaussian fit to the line profile is also shown:  the components are at 392\kms ,  25\kms FWHM  and 350\kms ,  50\kms FWHM.  }
\end{figure}
The line profile in a beam taking in the entire South-Eastern source is shown in Figure 9;  it combines a narrow ($\approx25$\kms\ FWHM) feature at 392\kms, close to the galactic velocity,  with a broader  blue component.  This is quite unlike the profile anywhere on the main source and confirms that this is region is independant of the supernebula.  It is discusssed further below. 
  \section{Discussion} 
  We have presented above  data showing that 1) the supernebula is not spherical but has arms or extensions, 2) a body of ionized gas including about 20\% of the total is blue-shifted by $\approx35$\kms\ relative to the main body of gas and is spatially offset from it, 3) the overlap of the blue shifted and main gas components has created an apparent but spurious velocity gradient, and 4) there is a weak pedestal of high-velocity gas from the supernebula. In addition we detect a new [SIV] and radio source $\approx20$ pc from the supernebula. 
\subsection{The Main Body of Gas and the Pedestal Emission}  
How can the embedded star cluster have created a nebula with these complex features?  First, what is the explanation of the  main body of gas with velocity less than  $\approx\pm50-60$\kms\ of the galactic?   Much of the ionized gas is at velocities that are probably gravitationally bound to the cluster.  For an estimated cluster mass of $10^6 M_\odot$ escape velocity at $R=2$pc is 65\kms~ and at $R=7$pc, corresponding to 0.5\arcsec on the sky, the escape velocity is 35\kms\ .  We see that only the high-velocity pedestal, $\approx10\%$  of the total ionized gas, is likely to escape the cluster from all radii.   Further, the velocity dispersion of the  bulk of the ionized gas is consistent with virial motion in the gravitational field of the embedded star cluster: following \citet{1988ApJ...333..821M} we find that the mass predicted from the observed FWHM for a radius of 1 pc and a $1/r^2$ density distribution is $4\times10^5 M_\odot$,  agreeing with the $5\pm2.5\times10^5 M_\odot$ found by other means \citep{2015Natur.519..331T}.    Unlike Galactic HII regions where stellar winds and pressure gradients determine the gas kinematics, gravitational forces in this source play the dominant role.  This has been seen in other extra-galactic embedded clusters \citep{2015ApJ...814...16B, 2012ApJ...755...59B} 

How can most of the cluster gas reach speeds much greater than the sound speed, but much less than standard 'cluster winds', which are typically several hundred \kms\, or the winds of massive stars which are even faster?   The recent work of \citet{2018MNRAS.478.5112S} and \citet{2017MNRAS.465.1375S} demonstrates that in very massive, compact clusters the metal-enriched winds of individual stars can dissipate their kinetic energy  in mutual shocks; in this situation the gas cools and is retained in the cluster and a fast wind does not develop.  The NGC 5253 star cluster appears in their models as the type example of this kind of cluster.  The star formation in the NGC 5253 cluster is extremely efficient in that most of the gas in the natal molecular cloud has been turned into stars \citep{2015Natur.519..331T}, but so compact that the activity of even this large population of stars does not expel the residual gas.   This model agrees well with the relatively low velocities of the bulk of the gas.   It further suggests that the high-velocity pedestal arises in winds from stars in the outer, less crowded regions of the cluster, and shows that only $\approx10\%$ of the gas twill escape from the cluster before being cooled.  
\subsection{The Arms and the Blue Feature: How are they Related}
Both the [SIV] and the radio continuum  'arms' extend N and W of the main source into the region dominated by the blue--shifted component of gas.  This suggests strongly that the 'arm' features are associated with the blue-shifted material. We now try to combine the spatial and kinematic distributions be combined into a full three-dimensional picture  of the region; there are several possible and plausible models. 
\begin{enumerate}
\item In the first model one central star cluster ionizes all the gas; the blue-shifted component, arms and spur are secondary features arising in interactions with the environment.  Mechanisms that could concievably create the observed structures:
\begin{enumerate}  
\item If a significant number of stars in the embedded cluster are driving outflows and their rotational axes are roughly aligned, the cluster may have created a bipolar outflow, of which we see only the blue-shifted side because of extinction. It is also possible for the outflow to  be intrinsically asymmetric, as reported by \citet{2013MNRAS.429.1747M}.  The two 'arms' observed could in this model be created by dense ambient gas blocking part of the flow. 
\item The blue-shifted gas may be expanding towards us into a region of low density material, a "champagne" flow, and the 'arm' structure created by interaction with ambient gas, as above.  This would agree with the observed structure of molecular gas in clouds southwest and northeast of the supernebula.
\item The embedded cluster may have created a shell of expanding gas on its NW side, whose edges are seen in the arms.  
\end{enumerate}
\item Another possibility is that the blue component of gas arises in a distinct body of ionizing stars, i.e. a  {\it sub-cluster}.  Sub-clusters are commonly found in Galactic clusters and it is thought that large clusters can grow by accreting them \citep{2014A&A...570A...2F}.  In this model the blue-shifted gas is mostly bound to a sub-cluster with central velocity $\approx-30$\kms\ relative to the main cluster.  The velocity offset between the main and blue-shifted gas features is consistent with rotation at distance less than $\simeq10$ pc.  In this picture the 'arms' could be gas escaping from or accreting onto the sub-cluster in the plane of the sky.  Since the hypothesized sub-cluster would be falling in from the far side of the supernebula it would be more highly obscured than the rest of the source; future observations
 could test this.  
\item The two types of models may coexist. For example, a sub-cluster may be present and gravitationally bound to the bulk of the blue gas, while the 'arms'  may represent a shell created by the main cluster, or accretion or expulsion episodes.   
\end{enumerate} 

Perhaps the strongest argument for the possibility of a blue-shifted sub-cluster is that the 'blue bump' has a roughly gaussian distribution around the central velocity.   This line profile would be
generated naturally by turbulent motion in a strong gravitational field, but not so readily in a pressure driven "champagne" outflow.  We can estimate the mass of stars in the hypothesized sub-cluster
from the FWHM of $\approx50$\kms\ and an assumed radius and mass distribution; if the radius is 1 pc and the distribution $1/r^2$ the total mass is $\approx 3\times10^5 M_\odot$.  This implies a sub-cluster roughly 1/3 the mass of the main emission source. This appears consistent with the rough 1/4 ratio of the [SIV] fluxes of the main and blue spectral features, to the limits of how well the two features can be separated.   Finally, it should be noted as arguing against this model that it has no candidate counterpart among the star clusters in the region, nor can it be associated with any feature in the molecular gas \citep{2017ApJ...846...73T,2017ApJ...850...54C}.

  \subsection{The South-East Source}
 The previously unreported [SIV] feature $\approx 1.5$\arcsec ($\approx 20$ pc) from the supernebula lies in a region of high optical extinction and obvious dust lanes.  It is undetected in the high resolution A array maps for which the largest detectable angular scale is  $\approx1.4$\arcsec\ , but does coincide  with a weak radio feature seen at 3.6 cm in an archival VLA map made with the B array and a 0.6\arcsec\ beam. This suggests  that the \HII~ region there is resolved out in the A array observations.  The line profile indicates that the bulk of the gas is flowing out of the source.  If we assume that the gaussian velocity feature at galactic velocity shows the component of gas which is gravitationally bound to the exciting stars,  the total stellar mass can be estimated (for a 1 pc radius and a $1/r^2$ density distribution) as $7.8\times10^4 M_\odot$. 
  
   \subsection{Summary and Conclusions}
We have presented measurements of the ionized gas in the great nebula in NGC 5253 with high spatial and spectral resolution.   The nebula is unexpectedly complex in both spatial structure and gas kinematics.  The images show that
\begin{enumerate}
\item  The ionized gas has has formed 'arms'  north-west of the main source.  
\item  The ionized gas appears to be ionization bounded on the southwest, towards CO cloud D4.
\item The line profiles show a distinct body of blue-shifted gas, offset from the main velocity by $\approx30$\kms\ and offset in space north-west  of the peak.  With the current spatial resolution we cannot determine the total extent affected by this feature.  We suggest that this blue feature creates the velocity gradient which has been interpreted as overall rotation of the nebula.  It could be an outflow  possibly related to the 'arms' in the same region. Alternatively,  the blue bump may be created in a sub-cluster about 1/3 as massive as the main source and bound to it gravitationally.  
\end{enumerate}
The youth of the system (the embedded star cluster(s) exciting the nebula are believed to be only $\approx1 Myr$ old)  suggests that the sub-structures we observe may be tracing the process by which the giant star cluster is assembled. Since the great bulk of the ionized gas does not have escape velocity from the star cluster it will be retained and if it can cool, it will continue to fuel more star formation.  It appears that star formation is not yet finished in this cluster.

\acknowledgements{Based on observations obtained at the Gemini Observatory, which is operated by the Association of Universities for Research in Astronomy, Inc., under a cooperative agreement with the NSF on behalf of the Gemini partnership: the National Science Foundation (United States), National Research Council (Canada), CONICYT (Chile), Ministerio de Ciencia, Tecnolog\'{i}a e Innovaci\'{o}n Productiva (Argentina), Minist\'{e}rio da Ci\^{e}ncia, Tecnologia e Inova\c{c}\~{a}o (Brazil), and Korea Astronomy and Space Science Institute (Republic of Korea), and on observations with the VLA, an  instrument of the National Radio Astronomy Observatory.  The National Radio Astronomy Observatory is a facility of the National Science Foundation operated under cooperative agreement by Associated Universities, Inc.  SCB thanks John Bally for useful discussions.  JT acknowledges support of NSF grant AST1515570. }

\bigskip
\facilities{VLA, Gemini:Gillett}

\software{AIPS, CASA, Ds9}
\bigskip

\noindent{\it Data Availability Statement}:  The data underlying this paper are available at the JVLA  and Gemini North Data Archives:  
https://archive.nrao.edu/archive/advquery.jsp and  https://archive.gemini.edu/.

\section{Appendix}
\begin{center}
\begin{figure}
\includegraphics[scale=0.4]{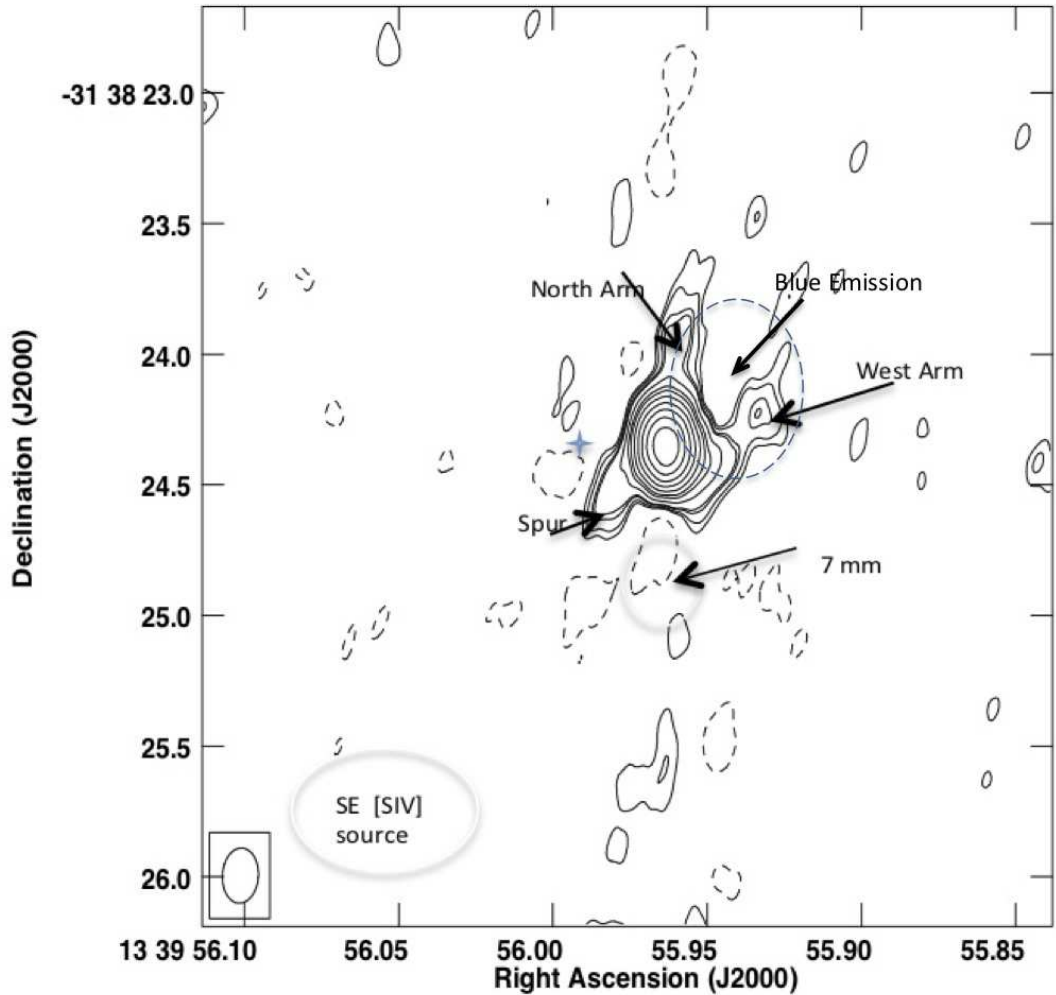}
\caption{A self-calibrated 33 Ghz map of NGC 5253, with beam in lower left.  Structures and sources discussed in the paper are marked on the figure. The 
north arm, west arm and south-east spur are marked with arrows. The approximate region of blue emission is marked by a blue dashed oval, and gray ovals show the locations of the south-east [SIV] source and the low-level 7mm emission detected by 
\citet{2004ApJ...602L..85T}.  The  small star shows the location of the secondary 2\um~ source found by \citet{2018ApJ...860...47C}. }
\end{figure}
\end{center}

\end{document}